\documentclass{article}
\usepackage{graphicx} % Required for inserting images
\usepackage{siunitx}
\usepackage{subcaption}
\usepackage{amsmath}

\newcommand*\C{\hspace{1mm} \mathrm{l}\hspace{-2mm}\mathrm{C}}
\newcommand*\R{\hspace{1mm} \mathrm{I}\!\mathrm{R}}

\newcommand*\re{\mathrm{re}}

\title{Monte Carlo simulation of random circuit sampling in quantum computing}
\author{Andreas Raab\footnote{Email: andreas.raab.mail@web.de}}
\date{\today}

\begin{document}

\maketitle

\begin{abstract}
We develop Monte Carlo methods for sampling random states and corresponding bit strings in qubit systems. To this end, we derive exact probability density functions that yield the Porter-Thomas distribution in the limit of large systems. We apply these functions in importance sampling algorithms and demonstrate efficiency for qubit systems with 70, 105, 1000, and more than one million ($2^{20}$) qubits. In particular, we simulate the output of recent quantum computations
without noise on a PC with minimal computational cost. 
I would therefore argue that random circuit sampling
can be conveniently performed on classical computers.
\end{abstract}


\section{Introduction}

Quantum computing is a rapidly evolving field in which tremendous progress has been made in recent years. A critical challenge for scaling
up quantum computers is, however, error control and error correction. 
Errors are typically inflicted by contact with the environment of the quantum device. 
This external "noise" leads to a loss of coherence in the quantum device so that its fidelity deteriorates. 

To benchmark fidelity and measure the performance of a quantum computer, 
random circuit sampling (RCS) with cross-entropy benchmarking (XEB) has been identified as a suitable task \cite{aru19,mor24,mor23}. 
RCS with a quantum computer can be divided into three steps: First, one samples a random circuit by classical means. one then runs the circuit on the quantum computer having a pre-defined initial state,
which typically is the ground state of the system. Finally, a bit string is read out from the
device, which effectively samples the bit string with a probability determined by the result state of
the calculation.

In this paper, we develop an alternative approach to RCS. The basic idea is to directly sample a random state of the underlying Hilbert space
instead of sampling a random circuit and applying it to an initial state. 
We show equivalence of both approaches in Sec. \ref{haar}, and we derive appropriate probability density functions for sampling random states in Sec. \ref{pdf}. 
In Sec. \ref{alg} we present the basic sampling algorithm, which we revise in Sec. \ref{large}
with respect to large numbers of qubits. Finally, we discuss our results in Sec. \ref{disc}.

\section{Random circuit sampling}

\subsection{Uniformly distributed quantum circuits}
\label{haar}

Let us consider a system with $n$ qubits. To establish appropriate sampling routines, we first identify the spaces and the sets from which to sample. 

The single-qubit Hilbert space is $\C^2$, so that the Hilbert space of the system is
 \[
 \mathcal{H} = (\C^2)^{\otimes n} = \C^D \quad (D = 2^n).
 \] 
Let us further assume a quantum calculation on the system with input $v_0$. In recent calculations, $v_0$ typically is the ground state of the system, but we note that for RCS the actual initial state does not matter. However, $v_0$ is a state in $\mathcal{H}$, i.e. $v_0 \in \mathcal{H} $, and 
$\| v_0 \| = 1$. A circuit in a quantum device is described by a unitary transform, $U$, and the corresponding quantum calculation basically applies 
$U$ to $v_0$, so that the (theoretical) output state is $v = Uv_0$. 

The unitary transforms on $\mathcal{H}$ are given by the unitary group $U(D)$, which is a subset of $\mathcal{H}^2 = \C^{2D}$. Since $U(D)$ is compact, 
there exists a unique norm-one Haar measure, $\nu$, which serves as the probability measure in RCS.
Due to translation-invariance, $\nu$ describes uniformly distributed circuits. 
For an observable, $F$, in RCS, the expectation value is given by
\begin{equation}
\label{rcs1}
\langle F \rangle = \int\limits_{U(D)} F(Uv_0) \, d\nu .
\end{equation}
Since $F$ is a function on the state space of $\mathcal{H}$, we argue in the following that we obtain the equivalent result by sampling 
uniformly distributed states:
\begin{equation}
\label{rcs2}
\langle F \rangle = \int\limits_{S(D)} F(v) \, d\mu .
\end{equation}
Here, $S(D)$ denotes the subset of states in $\mathcal{H}$,
\[
S(D) = \{ v \in \mathcal{H}: \, \|v\| = 1 \},
\]
and $\mu$ is the norm-one Lebesgue measure on $S(D)$.
We note that $\C^D$ is topologically isomorphic to $\R^{2D}$,
and that $S(D)$ is topologically isomorphic to the unit sphere,
$S_\re(2D)$, in $\R^{2D}$. In particular, the Lebesgue measure on $S(D)$ is invariant with respect to the action of $U(D)$.
\\[.3cm]
%
% see also https://www.ma.utexas.edu/users/gordanz/notes/measurable_spaces.pdf
%
{\bf Proposition:} The Haar measure $\nu$ uniquely determines $\mu$, and equations (\ref{rcs1}) and (\ref{rcs2}) are equivalent for an observable in RCS.
\\[.15cm]
{\em Proof}: 
We apply theorem 25 (p.384) of Ref. \cite{roy88} in the following: $U(D)$ is a
compact group which acts on the compact set $S(D)$. The group action is transitive,
since for all $v,w \in S(D)$ there exists a unitary transform $U$ such that
$v = Uw$. For each $v \in S(D)$ we further define the function
\[
\phi_v: U(D) \to S(D), \quad \phi_v(U) = Uv.
\]
These functions are continuous, so that for a compact subset $K \subset S(D)$,
$\phi_v^{-1}(K)$ is also compact. The group action is therefore proper, and we can apply the theorem. We note that the measure is given by 
$\mu = \nu \circ \phi_v^{-1}$ for any $v \in S(D)$ and that it 
also coincides with
the norm-one Lebesgue measure $\mu$.
$\rule{5pt}{5pt}$ \\[.3cm]
%
%
%See theorem 25 p.384 in Ref. \cite{roy88}
%

\subsection{Probability density functions}
\label{pdf}

In recent quantum calculations, the read-out after a calculation with $n$ qubits is a bit string of length $n$. 
The set of all possible bit strings, $\{ b \}$, provides a basis of $\mathcal{H}$, so that
a vector $v \in \mathcal{H}$ is given by
\[
v = \sum_b v_b b \, , \quad (v_b) \in \C^D.
\]
Each coordinate can further be represented by its real and imaginary parts, $v_b = x_b + i y_b$,
and the set of states is given by the unit sphere
\[
S_\re(2D) = \left\{ ((x_b,y_b)) \in \R^{2D}: \, \sum_b (x_b^2 + y_b^2 ) = 1 
\right\}.
\]
In order to sample a random state from $S_\re(2D)$, we first consider the
probability, $p_{b_1} = | v_{b_1} |^2$, of an output bit string $b_1$. We note that
if we sample or fix $v_{b_1}$ for a random state $v$, then the remaining part
is a random vector with norm $\sqrt{1-p_{b_1}}$,
\[
v^{(b_1)}= \sum_{b\neq b_1} v_b b\, \in \C^{D-1} \quad \left( \| v^{(b_1)} \|^2 = 1-p_{b_1} \right).
\]
This gives rise to a recursive pattern, which we exploit in the following.

The area of a sphere of radius $r$ in $\R^{2D}$, $S_\re(2D,r)$, is given by
\[
A(S_\re(2D,r)) = \frac{2 \pi^D}{(D-1)!} R^{2D-1}.
\]
Note that we use the notation $S_\re(2D) = S_\re(2D,1)$. Moreover,
\[
2\pi \int\limits_0^1dr \, \frac{ A(S_\re(2(D-1),r))}{A(S_\re(2D))} =
\int\limits_0^1 dr \, 2(D-1) r^{2D-3} = 1.
\]
We interpret 
\[
\tilde{\rho}_D(r) dr = 2(D-1) r^{2D-3} dr
\]
as the probability density function (PDF) of uniformly distributed states 
$v \in S_\re(2D)$ with $\| v^{(b_1)} \| = r$ for some bit string $b_1$. Since we
are mainly interested in probabilities in subsequent simulations, we switch to
the variable to $p = r^2$ and obtain the PDF
\[
\rho_D(p) dp = (D-1) p^{D-2} dp.
\]
We note that this also yields a PDF for $p_{b_1} = 1-p$, which
converges to the Porter-Thomas PDF for large
$D$:
\begin{eqnarray*}
(D-1) (1-p_{b_1})^{D-2} \, dp_{b_1} & = & \frac{D-1}{D-2} \left( 1- \frac{\bar{p}_{b_1}}{D-2} \right)^{D-2} d\bar{p}_{b_1} \quad ( \bar{p}_{b_1} = p_{b_1} (D-2)) \, , \\
& \approx & 
e^{-\bar{p}_{b_1}} \, d\bar{p}_{b_1} \\
& = &  D e^{-D \tilde{p}_{b_1}} \, d\tilde{p}_{b_1}\, \quad \left( \tilde{p}_{b_1}
= \frac{\bar{p}_{b_1}}{D} = \frac{D-2}{D} p_{b_1}  \right).
\end{eqnarray*}
However, we do not study the technical details of this limit here, since we continue working with the exact PDF, $\rho_D$. 

\subsection{Sampling algorithm}
\label{alg}

The Monte Carlo methods developed in this paper make use of importance sampling. 
Let $x \in [0,1]$ be a random variable, let $f$ be a continuous positive PDF, and let
\[
x = F(y) = \int\limits_a^y f(z) \, dz \quad (-\infty \leq a \leq y \leq b \leq \infty).
\]
Since $f$ is positive, $F$ is bijective so that $y = F^{-1}(x)$ is well-defined.
Moreover,
\[
\frac{dx}{dy} = f(y),
\]
and we obtain for an integrable function (observable), $g$,
\[
\langle g \rangle = 
\int\limits_a^b f(y) g(y) \, dy = \int\limits _0^1 dx \, g(F^{-1}(x)).
\]
We note that $dx$ is the image measure of $f(y)dy$. For the PDF $f(p) = \rho_D(p)$ we obtain in particular
\[
F(p_b) = \int\limits_0^{p_b} (D-1) p^{D-2} \, dp = p_b^{D-1} \, , 
\quad F^{-1}(x) = x^{\frac{1}{D-1}} \, .
\]

To sample a random state, we start with sampling the bit-string probabilities,
$p_b$. As mentioned in Sec. \ref{pdf}, we can perform this recursively: \\[.3cm]

\begin{figure}[h]

\begin{subfigure}{0.5\textwidth}
\includegraphics[width=0.9\linewidth, height=3cm]{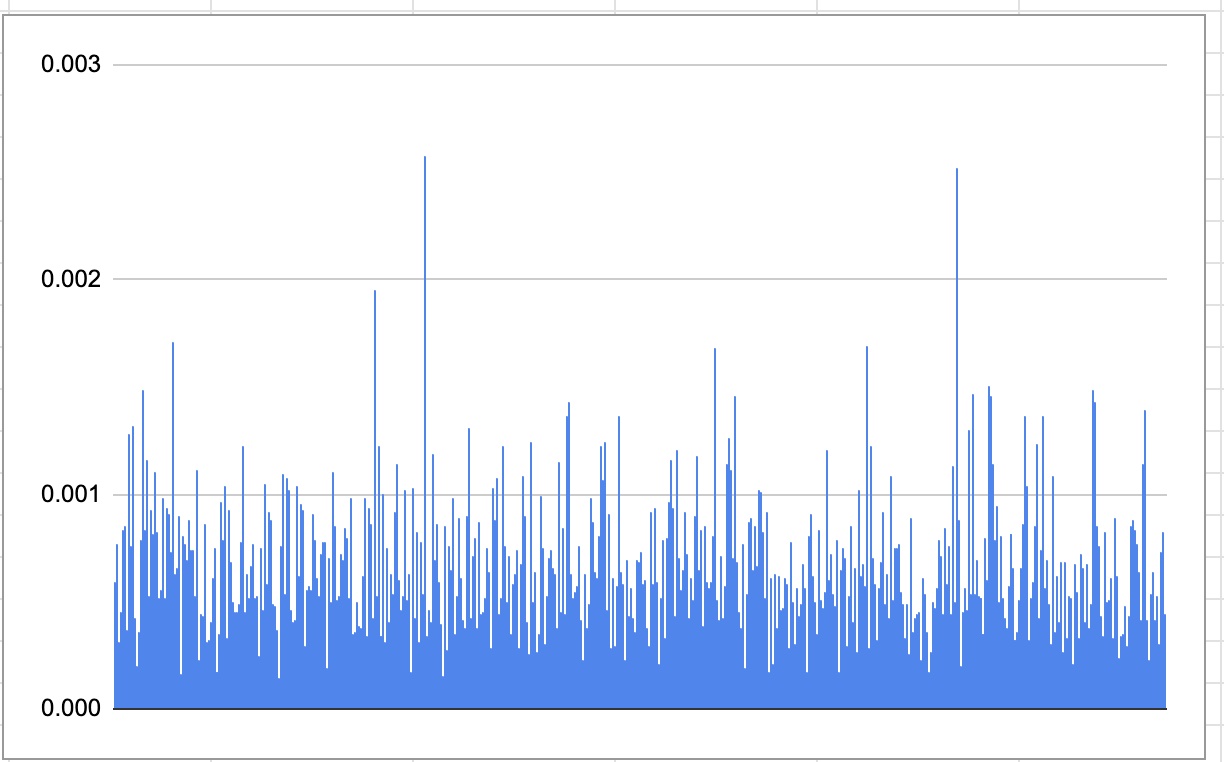} 
\caption{Monte Carlo simulation}
\label{fig:rcs_12_mc}
\end{subfigure}
\begin{subfigure}{0.5\textwidth}
\includegraphics[width=0.9\linewidth, height=3cm]{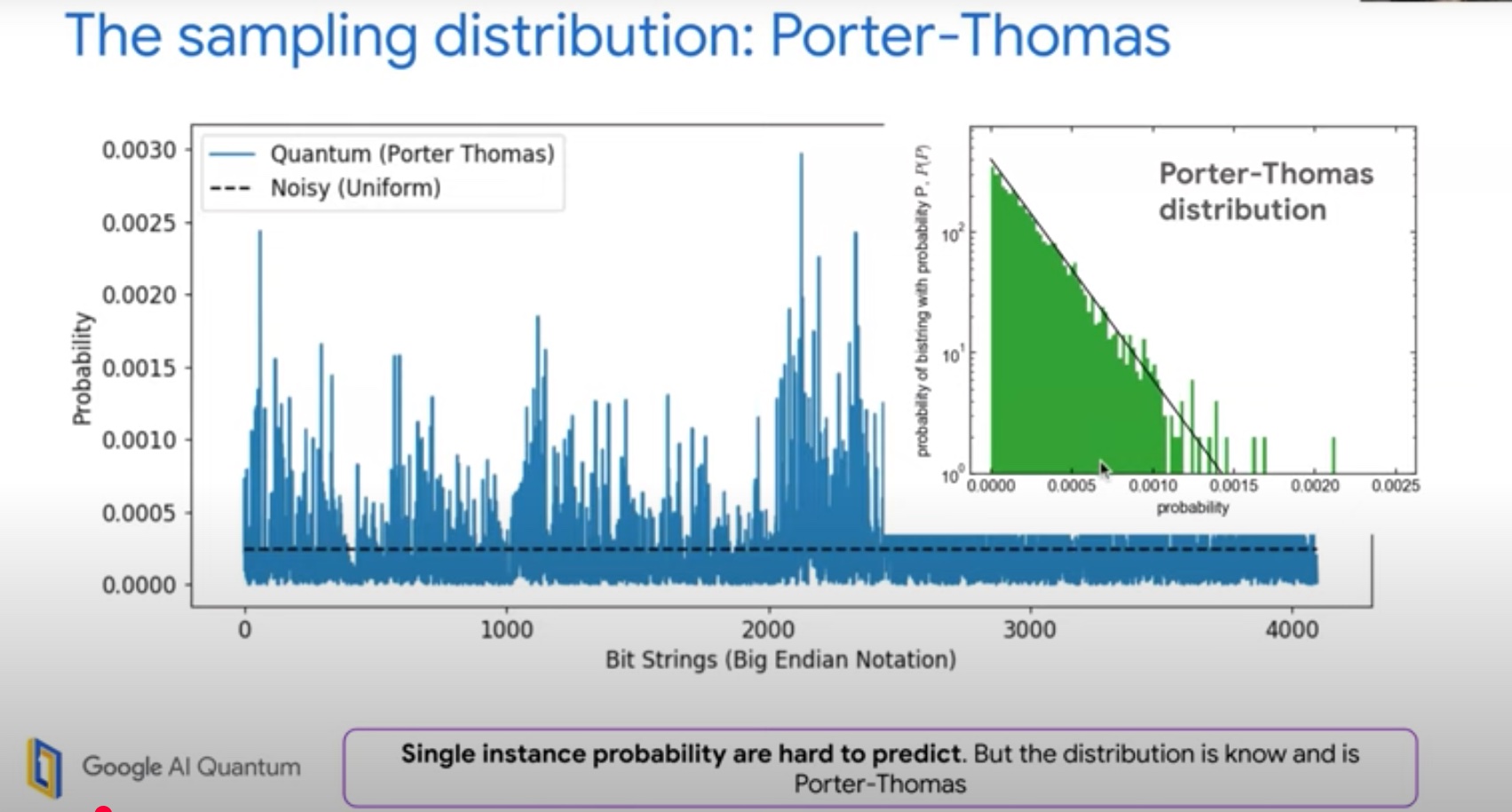}
\caption{Example in Ref. \cite{mor23}}
\label{fig:rcs_12_qc}
\end{subfigure}

\caption{Component probabilities of a random state of a system with 12 qubits}
\label{fig:rcs_12}
\end{figure}

{\bf \underline{Algorithm 1}} \\
\begin{enumerate}
    \item Sample $x_1 \in [0,1]$, and set $p_1= 1 - x_1^{\frac{1}{D-1}} $
    \item For $2 \leq k \leq D-1$: Sample $x_k \in [0,1]$, and set 
\[
        p_k= \left( 1 - \sum_{l=1}^{k-1} p_l \right) 
        \left( 1 - x_k^{\frac{1}{D-k}} \right)
\]
\item Set
\[
        p_D= \left( 1 - \sum_{l=1}^{D-1} p_l \right) 
\]
\item Sample $D$ phase angles in $[0,2\pi)$
\item Sample a permutation $\sigma$ of $(1,2,...,D)$ \\
\end{enumerate}
We note that the last step reflects the fact that for any state $v = \sum_b v_b b$ and
any permutation $\sigma$, $v_\sigma = \sum_b v_{\sigma(b)} b$ is also a state.
In particular, performing steps 1 to 4 yields an equivalence class of states,
$[(v_{\sigma(b)})]$, and in the last step we sample an element in that equivalence class.
However, we illustrate this algorithm for a system with 12 qubits in Fig. \ref{fig:rcs_12_mc}, which is
analogous to the example presented in Ref. \cite{mor23}. The algorithm is implemented
in a spreadsheet without step 4, because only probabilities, $p_b$, are depicted (see also Ref. \cite{raa25}).

\subsection{Sampling states of larger systems}
\label{large}

\begin{figure}
    \centering
\includegraphics[width=0.8\linewidth]{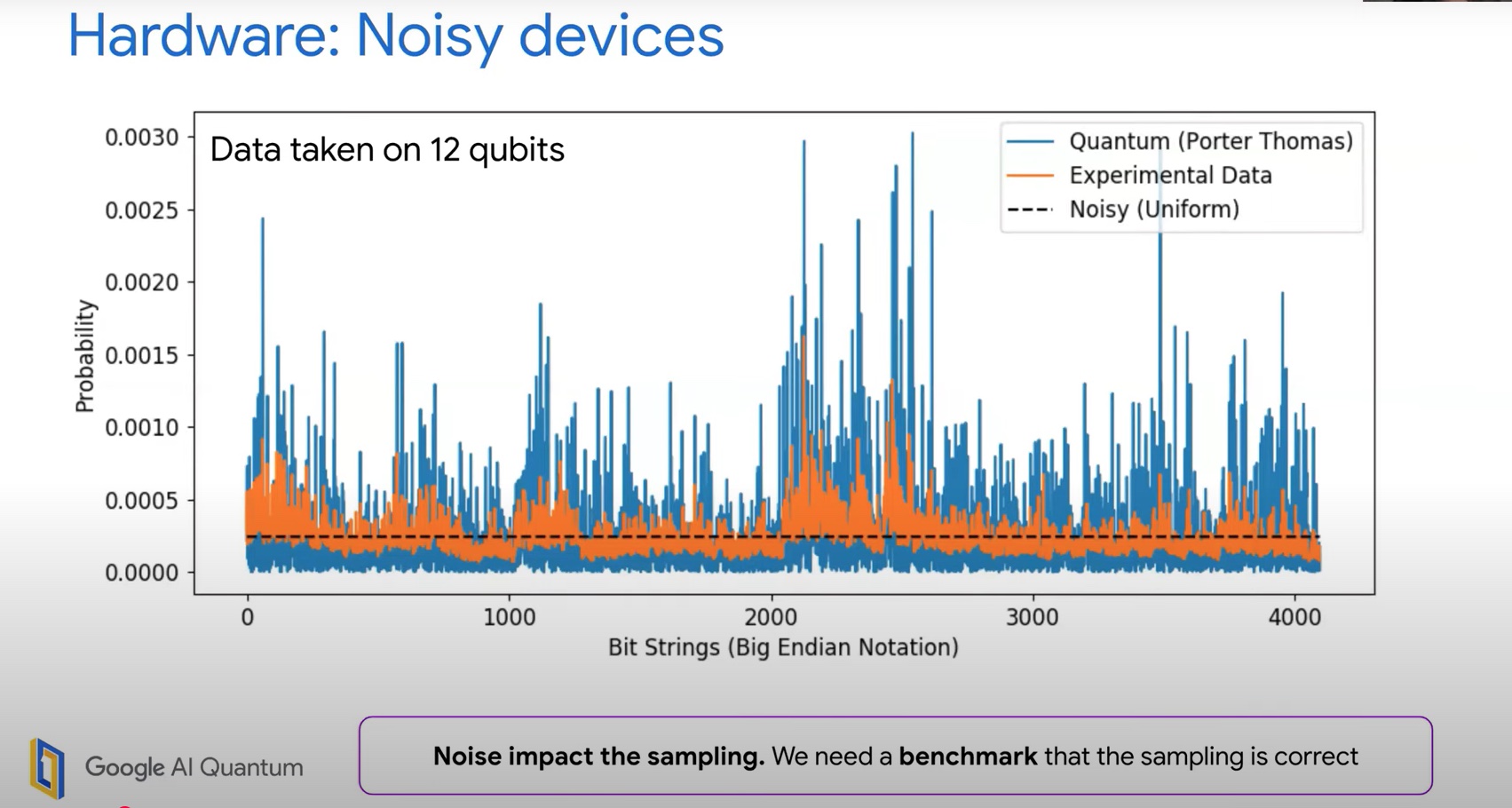}
    \caption{Example in Ref. \cite{mor23}: Component probabilities of a random state of a system with 12 qubits. Blue bars depict the values of an exact quantum calculation on a classical device, while orange bars depict experimental (statistical) values obtained with a quantum device.}
    \label{fig:rcs_2_qc}
\end{figure}

For larger systems, fully capturing a random state quickly becomes infeasible. 
The reason is that reading out the result of a quantum computation is a measurement
process, which leads to a collapse of the wave function
according to the Copenhagen interpretation of quantum mechanics. In particular,
reading out the result yields a bit string, $b$, with probability $p_b = |v_b|^2$, where
$v_b$ is the corresponding component of the result state $v = \sum_b v_b b$.
The result state $v$ is therefore only statistically accessible and can
only be evaluated using expectation values of observables, which are obtained
by repeating the same computation many times and taking statistical averages. 
An example is shown for component probabilities of a random state in Fig. 
\ref{fig:rcs_2_qc}.
We therefore sample only one component when simulating larger systems, which is sufficient for
comparison with results of recent quantum computations, see Ref. \cite{aru19,mor24,goo25,gao25}. 

A benchmark observable in quantum RCS is cross-entropy fidelity, $F_{\mbox{\tiny XEB}}$, which is calculated as follows: 
One first samples a random state, $v$, and by reading the output bit string, $b$,
one samples $b$ from $v$ with probability $p_b$. Determining further $p^{(sim)}_b$ with an accompanying classical simulation one calculates
\[
F_{\mbox{\tiny XEB}} = D \left\langle p^{(sim)}_b \right\rangle - 1 \, .
\]
From a theoretical perspective, the expectation value of this two-step sampling process 
is given by
\[
 F_{\mbox{\tiny XEB}} =   D \!\!\!\int\limits_{S(D)} \sum_b p_b^2  \, d\mu  \, - 1
\]
Note that $d\mu$ is the PDF for sampling a random state and that $p_b$ is the discrete 
probability for sampling $b$ during read out.
Since $\mu$ is invariant with respect to permutations of indices (which are unitary transforms), the integral reduces to
\[
 F_{\mbox{\tiny XEB}} = D^2 \!\!\!\int\limits_{S(D)} p_{b_0}^2  \, d\mu - 1= 
 D^2 \! \int\limits_0^1 (1-p)^2 \rho_D(p) \, dp \, - 1
 = \frac{D-1}{D+1}\, ,
\]
where $b_0$ is a fixed but otherwise arbitrary coordinate (bit string).

To achieve numerical stability with increasing number of qubits,
we rescale probabilities by a factor of $D$,
i.e. we introduce
$\bar{p} = pD$ in the sampling algorithm, and we 
rewrite the inverse distribution function as follows  
\[
\bar{p} = D p = D \left( 1 - x^{\frac{1}{D-1}} \right) = D \left( 1 - \exp\left( 
\frac{\ln{x}}{D-1}\right)\right) = - D \,\, \mbox{expm1} \left( 
\frac{\ln{x}}{D-1}\right)
\, .
\]
We note in particular that for $D \to \infty$ we obtain
\[
D \left( 1 - \exp\left( 
\frac{\ln{x}}{D-1}\right)\right) 
\to \left. - \frac{d}{dt} \exp(t \ln{x}) \right|_{t=0} = - \ln{x} \, ,
\]
i.e., for large $D$, we basically sample from the Porter-Thomas PDF,
$\exp(-\bar{p}) d\bar{p}$. The revised algorithm for large systems with $n$ qubits 
($n \geq 45$) now reads as follows: \\[.3cm]

{\bf \underline{Algorithm 2}} \\
\begin{enumerate}
    \item Sample $x \in [0,1]$. For $n \leq 1000$ ($D = 2^n$), set
\[
\bar{p} = - D \,\, \mbox{expm1} \left(  \frac{\ln{x}}{D-1}\right), 
\]
else $\bar{p} = - \ln{x}$
\item Sample a random bit string of length $n$
\item Repeating steps 1 and 2 $N$ times, $F_{\mbox{\tiny XEB}}$ is estimated as
\[
\bar{F}_{\mbox{\tiny XEB}} = \frac{1}{N}\sum_{k=1}^N \bar{p}_k^2  - 1.
\]    
\end{enumerate}
Note that convergence of the estimate to its theoretical value is guaranteed 
by the central limit theorem. We also note that sampling a bit string of length $n$ is
the time-consuming step for large systems, which can however be improved by parallelization. 

For test purposes, algorithm 2 was implemented on a PC, and the results
are summarized in table \ref{MC_sim_data}. Qubit systems with 70, 105, 
1000, and 1048576 ($2^{20}$) qubits are chosen for test runs.

% Requires: \usepackage{siunitx}
\begin{table}[h]
    \centering
    \begin{tabular}{l l l l l}
        \hline
        Qubits & Samples & $F_{\mbox{\tiny XEB}}$  & Time \\
        \hline
        70 & \num{E7} & $0.999 \pm 0.0043$ & 1.0s \\
        105 & \num{E7} & $0.999 \pm 0.0043$ & 1.2s \\
        1000 & \num{E7} & $0.999 \pm 0.0043$ & 7.2s \\
        1048576 & \num{E5} & $ 1.02 \pm 0.043$ & 95s \\
        \hline
    \end{tabular}
    \caption{Monte Carlo simulation of quantum RCS experiments without external noise. $F_{\mbox{\tiny XEB}}$ 
    is calculated with a 3$\sigma$ error estimate using algorithm 2 on a
    3.1 GHz Dual-Core Intel Core i5 CPU. Note that the 70 qubit example
    and the 105 qubit example refer to experiments in Refs. \cite{mor24,goo25,gao25},
    while the examples with 1000 and 1048576 ($2^{20}$) qubits
    refer to expected experiments on future quantum computers.}
    \label{MC_sim_data}
\end{table}

%\newpage

\section{Discussion}
\label{disc}

Random circuit sampling has become a well-established task for measuring
fidelity and performance in quantum computing. 
The quantum problem is hard to solve on a 
classical device since the computational effort increases exponentially 
with the number of qubits. In particular, running a random circuit on a 
quantum computer typically yields a highly entangled result state. Due to 
the high degree of entanglement, 
the full Hilbert space needs to be considered, so that both memory 
consumption and run times increase 
exponentially for classical devices. Therefore, quantum computers
can provide results, where the number of qubits makes calculations
on classical devices impossible. 

However, the result state of a calculation on a quantum computer is only 
statistically available, since reading the output is a 
measurement process, which leads to a collapse of the wavefunction.
We can in principle only obtain 
expectation values of observables with quantum computers, which are 
calculated by repeating the same computation many times and 
taking statistical averages. In particular, the bit strings read out in quantum
RCS are basically sampled with a probability that is given by the probabilities of the components of the result state, i.e., the squared absolute values of its components.
We note also that to verify the correct probability
distributions, auxiliary classical calculations are usually required.

In this paper, we develop an alternative approach to RCS. Instead of
applying a random circuit to an initial state of a qubit system,
we directly sample components of a random state. We have shown equivalence 
of both approaches in Sec. \ref{haar},
which means that the result of applying a random circuit to an initial
state of a qubit system is a random state. To perform the task of
RCS, quantum calculations (i. e. using random circuits) are therefore not strictly required.

In Sec. \ref{pdf}, we derive
exact probability density functions for sampling random states, which
we use in sampling algorithms in Sec. \ref{alg} and in Sec. \ref{large}.
Test runs on a PC demonstrate the extremely low computational cost of
this approach; see table \ref{MC_sim_data}, so that I would argue that
the task of RCS can
be conveniently performed on classical PCs.

\end{document}